\documentclass[aps,pre,twocolumn,showkeys,10pt]{revtex4-1}
\usepackage{amssymb}
\usepackage{amsmath}
\usepackage{graphicx}% Include figure files
\usepackage{subfigure}
\usepackage{multirow}
\usepackage{bm}% bold math
\usepackage{color}%
\begin{document}

%\preprint{AIP/123-QED}

\title{The electronic and optical  properties of graphene nanoribbons under the influence of the periodic strain}% Force line breaks with \\
%\thanks{Footnote to title of article.}
\author{Chunwen Zhang$^1$, and W.X. Yan$^{2 *}$}
\affiliation{ $^1$Insititute of Theoretical Physics, Shanxi University, Taiyuan, 030006, People's Republic of China
\\ $^2$School of Physics and Electronic Engineering, Shanxi University, Taiyuan, 030006, People's Republic of China}
\email{wxyansxu@gmail.com}
%\affiliation{ }

\begin{abstract}
The electronic and optical properties of graphene nanoribbons under uniaxial periodic
strain have been explored using various nearest-neighbor hopping patterns. It is found
that by properly selecting hopping patterns, momentum-resolved gaps within
minibands emerge, modifying the energy band structure to exhibit hollowed-out
profiles, and enhancing peak intensity in local density of states but reducing peak
count. The optical transitions are impacted by altered parity symmetry of
wavefunctions, causing changes in optical selection rules. The parity of wavefunctions for strained GNRs has been established through 
 rigorous mathematical proof, whereby the optical selection rule is determined for the strained GNRs. The absorption curves arise from a
complex interplay between diminished velocity matrix elements and escalated joint
density of states.
\end{abstract}

\pacs{Valid PACS appear here}% PACS, the Physics and Astronomy
                             % Classification Scheme.
%Use showkeys class option if keyword
                              %display desired
\maketitle

\section{Introduction}

Graphene, an exceptional material comprising a single layer of carbon atoms arranged in a hexagonal lattice, possesses remarkable properties such as high electrical conductivity, unparalleled mechanical strength, and superior thermal conductivity\cite{graphene_ele1,graphene_ele2,marconcini_rev}. When graphene is engineered into graphene nanoribbons (GNRs), quantum confinement and edge effects emerge, imparting novel properties absent in its bulk case\cite{marconcini_rev,wang_elerev,kumar_rev,louie_ele,louie_ele2,zhanggpcpl}. The electronic behavior of GNRs is profoundly influenced by external fields\cite{cresti08,linm,tejada_ele,wei_ele,ooi_opt,almeida22,xia_opt,chenjun,chnafassc}, chemical doping\cite{chemdoping0,chemdoping1,chemdoping2}, and strain\cite{pereira_strain,vozmed_strain,chauwin_strain,Naumis111,wang_strain,naumis14,Yanw,chnafa_strain}, offering avenues for tailored functionalities. 
The distinctive optical properties of GNRs hold immense promise for diverse optoelectronic applications, underpinned by their tunable bandgap, robust light-matter interactions, and intriguing nonlinear optical properties\cite{hsu07,liao_opt,prezzi_opt,sasaki,mflin11,saroka17,shukla2,gundra11,sasaki20,monozon22_opt,bang_opt,cresti_opt,saroka19_opt}. These features unlock new horizons in the design and development of advanced photonic and optoelectronic devices.

Recent advancements in the bottom-up synthesis of GNRs have facilitated atomic-precision control over virtually all structural parameters\cite{nature18a,nature18b,jpc21}, achieved through the deliberate design and self-assembly of small-molecule precursors. This achievement positions GNRs as an alluring platform for exploring topological phases, as graphene's transition to a semiconductor, induced by lateral confinement with specific edge geometries, opens doors to unprecedented opportunities in materials science and nanotechnology.

In the tight binding framework and Keldysh-Green's formalism, Cresti and  collaborators delved into the electronic structure and transport characteristics of Dirac particles in GNRs\cite{cresti08}, both in the absence and presence of a magnetic field (through Peierls substitution). Their work illuminated the chiral behavior of the Dirac particles in GNRs, which was found to be modulated by external gate potentials. Meanwhile, Naumis and coauthors mapped zigzag graphene sheet tuned by oscillating strain along the armchair direction into an effective tight binding chain whose unitary cell consists of four nonequivalent carbon atoms in a periodic manner\cite{naumis14}. They numerically investigated the influence of this strain on the energy spectrum and density of states (DOS). Subsequently, the same research group expanded their investigation to GNRs subjected to similar uniaxial strain profiles, numerically calculating the Landauer-B\"uttiker conductivity by employing scattering matrix computations\cite{Naumis111}.

Hsu and Reichl conducted a study on the interband optical transition properties of pristine GNRs\cite{hsu07}, numerically contrasting the optical selection rules between GNRs and carbon nanotubes. Shortly afterwards,  Sasaki et al. and Chung et al. independently derived analytical expressions for the optical selection rules of pristine GNRs, which were validated through numerical simulations of interband optical absorption\cite{sasaki}. Saroka et al. further broadened the investigation by comprehensively comparing the electronic and optical properties of GNRs and carbon nanotubes, extending their analysis beyond the single-k paradigm using the transfer matrix method and a dispersive tri-diagonal effective Hamiltonian. This comprehensive study emphasized the connection between interband and intraband selection rules and the parity symmetry of (envelope) wavefunctions, particularly highlighting the transformation between edge and bulk states through the analysis  of wavefunction symmetry\cite{saroka17}.

Inspired by these works, the present study aims to incorporating the strain into GNRs, and study the influence of the strain on electronic and optical properties of GNRs. In order to study the strain effect on the GNRs in a tractable  manner, several hopping patterns due to strain have been designed to enable analytical study on the electronic structure minibands as well as the corresponding interband and intraband optical properties. 
The electronic minibands and local density of states (LDOS) has been compared and discussed under different hopping patterns.
The optical properties has been investigated through studying the intraband and interband velocity operator matrix,
and the relationship between the interband and intraband optical selection rules and the  concomitant  symmetry breaking and restoration of the wavefunction has been established.

The work is structured as follows: In Section II, the model is mapped onto a dispersive tri-diagonal effective Hamiltonian, enabling the derivation of analytical results for the electronic structure and wavefunctions using the transfer matrix. Section III presents numerical findings for minibands, local density of states (LDOS), and wavefunctions. Section IV delves into intraband and interband velocity matrix elements, exploring how strain modifies the optical selection rules, and also discusses the resulting absorption properties. The concluding remarks is presented in section V.
\begin{widetext}

\begin{figure}[htb!] 
\centering 
\subfigure[]{ 
\label{Fig01a} 
\includegraphics[height=5.cm,width=0.6\textwidth]{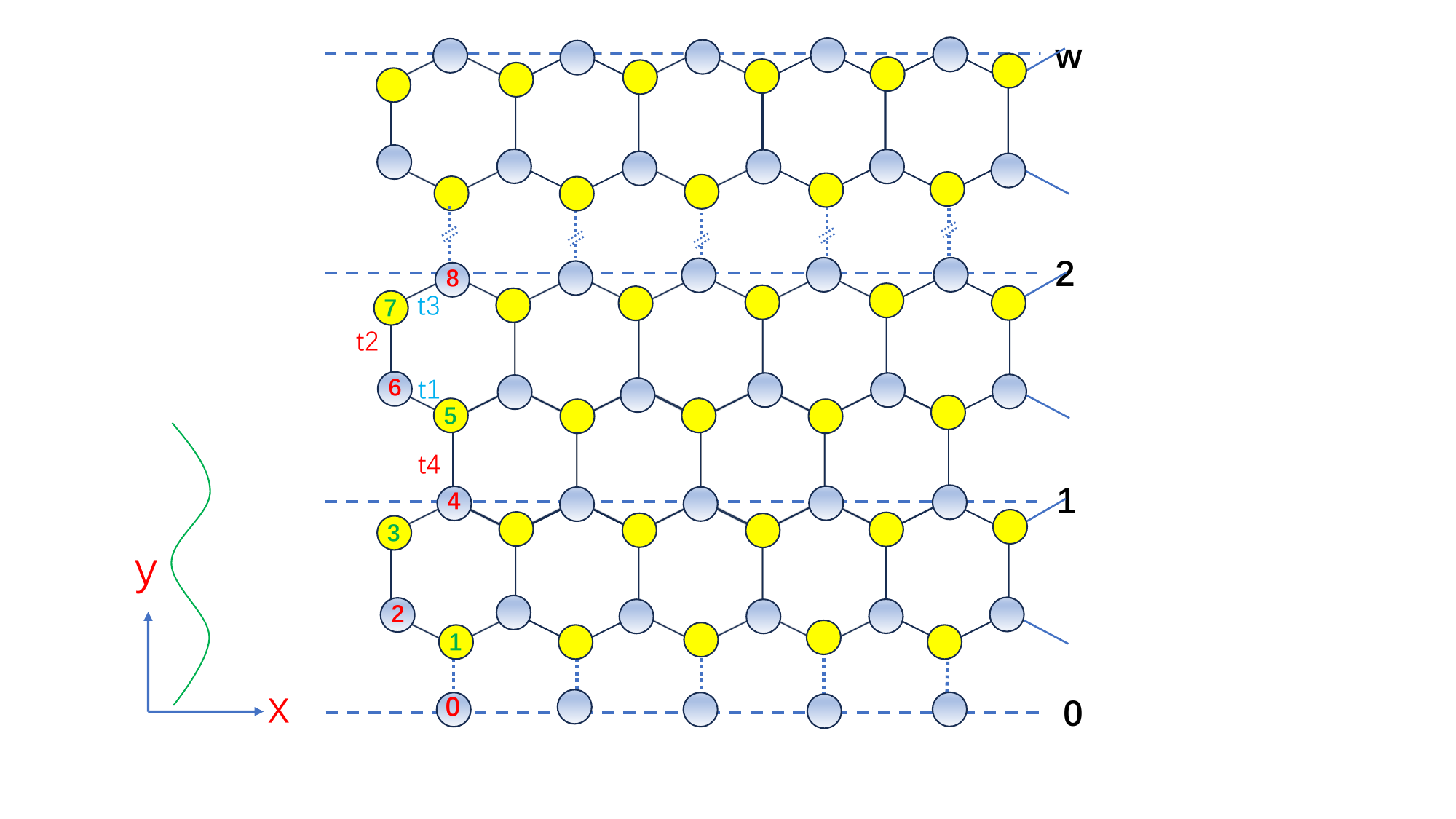}}
\subfigure[]{ 
\label{Fig01b} 
\includegraphics[width=0.35\textwidth]{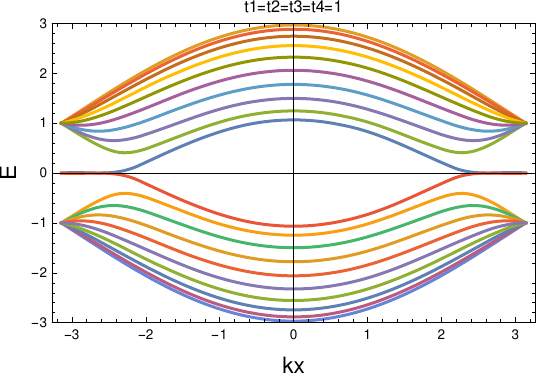}}  
\caption{(a) The schematic diagram of GNRs with four non-equivalent atom arranged in periodic manner in the unit cell where $t_i, (i=1,2,3,4)$ are the hopping parameters between the neighboring carbon atoms, the total numer of atoms in unit cell is $N=4w$. (b) The energy minibands of the pristine GNRs whose unit cell contains 20 carbon atoms, all hopping terms are the same, and assumed to be of unity ($t=1.0$) hereafter.} 
\label{Fig01}
\end{figure}
\section{The model and analytical results}
For the GNRs with arbitrary length $L_x=N_x a$ along the $x-$direction,  the Bloch wave function can  be written in the tight-binding framework as follows\cite{cresti08},   
\begin{eqnarray}
|\psi_j\left(k_x,\bm{r}\right)\rangle=\left(\frac{1}{\sqrt{N_x}}\right)\sum\limits_{\bm{R}_m}\exp\left[i\bm{k}\cdot\left(\bm{R}_m+\bm{d}_j\right)\right]\left| \phi_z(\bm{r}-\bm{d}_j-{\bm R}_{m})\right\rangle 
\end{eqnarray}
\end{widetext}
\noindent where $\bm{k}=\left(k_x,0\right),\bm{R}_m=\left(ma,0\right)$ is the periodic ribbon vector along the zigzag ($x-$) direction, $\bm{d}_j=(d_{jx},d_{jy})$ is the position vector of the $j-$th carbon atoms in the $i-$th unit cell, 
and $\phi_z$ is the carbon atomic $\pi$-orbital.
In order to study the strain effect on electronic and optical properties of the GNRs, the applied strain is assumed to modify the hopping term in an periodic manner, that is to say, 
the hopping terms between the nearest neighboring carbon atoms are changed from  the uniform $t$ to $t_1,t_2,\cdots,t_j$ along the direction $y$.
In this work, we focus on GNRs whose unit cell consists of periodic four nonequivalent carbon atoms and the nearest hoppings $(t_i, i=1,2,3,4)$ are designed to be different in contrast to the pristine GNRs.
As shown in Fig.\ref{Fig01a}, the number of zigzag chains containing the nonequivalent four carbon atoms in a periodic manner is $w$, and then the total number of carbon atoms in the unit cell is $N=4w$.

 In the above assumed scenario, the effective dispersive Hamiltonian can be written as follows\cite{cresti08,saroka17,Naumis111}
\begin{eqnarray}
\label{ham}
%\begin{aligned}
H=\begin{pmatrix}
0 & h_1 & 0 & 0 & 0 & 0 \\
h_1 & 0 & h_2 & 0 & 0 & 0 \\
0 & h_2 & 0 & h_3 & 0 & 0 \\
0 & 0 & h_3 & 0 & h_4 & 0  \\
0 & 0 & 0 & h_4 & 0 & h_1 \\
0 & 0 & 0 & 0 & h_1 & 0 \\
\end{pmatrix}
%\end{aligned}
\end{eqnarray}
where $h_j=qt_j$ ($q=2\cos(k_x a/2)$, and $a=1$ is assumed hereafter), when $j$ is an odd number, and $h_j=t_j$, otherwise.
Then Schr\"odinger equation can be derived in the tight-binding representation as follows, 
\begin{eqnarray}
\label{eq3}
\begin{aligned}
 & h_4 c_{j-1}-c_jE + h_1 c_{j+1}   =0; \quad j=4p-3 \\
& h_1 c_{j-1} -c_jE + h_2 c_{j+1} =0,  \quad j=4p-2\\
&h_2 c_{j-1} -c_jE + h_3 c_{j+1} =0,  \quad j=4p-1\\
& h_3 c_{j-1} -c_j E + h_4 c_{j+1} =0, \quad j=4p
\end{aligned}
\end{eqnarray}
where $p=1,{\ldots}{\ldots}, w$, $w=N/4$ (see Fig.1).
The above equation can be rewritten as the transfer matrix form:
\begin{eqnarray}
\label{eq5trans}
\begin{aligned}
& \begin{pmatrix}
c_{4(p-1)+i}\\
c_{4(p-1)+i+1}
\end{pmatrix}=T_i\begin{pmatrix}
c_{4(p-1)+i-1}\\
c_{4(p-1)+i}
\end{pmatrix}
\end{aligned}
\end{eqnarray}
\vspace{0.3cm}
where the transfer matrix $T_i~ (i=1,2,3,4)$ is defined as follows:
\onecolumngrid

\begin{eqnarray}
\begin{aligned}
 & T_1=\begin{pmatrix}
0 & 1\\
-\frac {h_4}{h_1} & \frac{E}{h_1}
\end{pmatrix}, \quad T_2=\begin{pmatrix}
0 & 1\\
-\frac{h_1}{h_2} & \frac{E}{h_2}
\end{pmatrix}, \quad 
T_3=\begin{pmatrix}
0 & 1\\
-\frac{h_2}{h_3} & \frac{E}{h_3}
\end{pmatrix}, \quad T_4=\begin{pmatrix}
0 & 1\\
-\frac{h_3}{h_4} & \frac{E}{h_4}
\end{pmatrix}
\end{aligned}
\end{eqnarray}
Then the total transfer matrix  ${\mathcal T}$ can be defined as follows,
\begin{eqnarray}
\label{eq6}
\begin{aligned}
& {\mathcal T}=\prod\limits_{i=1}^4 T_i=\frac1{h_1h_2h_3}\begin{pmatrix}
(-E^2+h_2^2)h_4 & E(E^2-h_1^2-h_2^2)\\
E(-E^2+h_2^2+h_3^2) & \left[E^4+h_1^2h_3^2-E^2(h_2^2+h_1^2+h_3^2)\right]/h_4 
\end{pmatrix} 
\end{aligned}
\end{eqnarray}

\twocolumngrid
The total transfer matrix ${\mathcal T} $ can be diagonalized through the similarity transformation: $S{\mathcal {T}}S^{-1}=(\lambda_1,\lambda_2)^T$, and the eigenvalues  can be obtained as, 
$(\lambda_1,\lambda_2)^T=-(\exp{- i\theta}, \exp{+ i\theta})^T$, 
where $\theta$ is defined through the relation: $\cos\theta=-({\mathcal T}_{11}+{\mathcal T}_{22})/2$, and the corresponding similarity transformation matrix $S$ is, 
\begin{eqnarray}
\label{eq13}
S=\begin{pmatrix}
1 & 1\\
\xi_1 & \xi_2
\end{pmatrix},
\end{eqnarray}
where 
\begin{eqnarray}
\label{eq12}
\xi_{1,2}=\frac{\lambda_{1,2}{\mathcal T}_{21}}{\lambda_{1,2}{\mathcal T}_{11}-1}~.
\end{eqnarray}
For GNRs unit cell shown in Fig.1, inserting Eq.\eqref{eq13},\eqref{eq12} into ${\mathcal T}^{{w}}=S \lambda^{w}S^{-1}$\cite{saroka17}, and applying the boundary condition for the transmission, i.e., $\left(T^w\right)_{22}=0$, one can obtain,
\begin{eqnarray}
\label{qtheta}
{\mathcal T}_{11}\sin w\theta +\sin \left(w+1\right)\theta =0
\end{eqnarray}
Solving the above equation, the momentum $q$ and energy $E$ can be obtained as follows, 
\begin{widetext}
\begin{eqnarray}
\label{eqnum}
\begin{aligned}
& q=
\left[{-\frac{t_2t_4 \bigg(\left(\alpha_2-2\alpha_1{\cos}\theta \right) \sin w\theta +\left(\alpha_1-\alpha_3\right) \sin (w+1)\theta \bigg)\sin w\theta}{t_1t_3\bigg(-\alpha_1\sin^2 w\theta+\alpha_2\sin w\theta \sin\left(w+1\right)\theta -\alpha_3 \sin^2(w+1)\theta\bigg)}}\right]^{1/2} \\
& E=
\pm t_2 \left|\csc w\theta\sin\theta \right|\left[-\frac{\alpha_1}{-\alpha_1+\alpha_2\csc w\theta \sin(w+1)\theta -\alpha_3\csc^2 w\theta\sin^2(w+1)\theta}\right]^{1/2}
\end{aligned}
\end{eqnarray}
where $\alpha_1=t_1t_3 t^2_4, \alpha_2=t_2t_4(t^2_1+t^2_3),\alpha_3=t_1 t^2_2 t_3$, and the symbol $\pm$ represents the conduction (+) and valence (-) bands respectively.

Form the above equation, it can be clearly seen that both $q(\theta)$ and $E(\theta)$ remains unchanged under the transform $\theta\rightarrow -\theta$.

\section{Electronic Properties of GNRs due to strain}

Many groups had studied the energy gaps of $N$-A(Z)GNRs' under the influence of uniaxial strain\cite{lunanores,linanores,kliros,liaoepjb}. 
Their results shared a common feature that the changing trends of these energy gaps are characterized by the number of carbon atoms $(N)$ in the unit cell of GNRs, 
which is classified as three families of $3p, 3p+1, 3p+2$ (where $p$ is an integer)\cite{lunanores,linanores,kliros,liaoepjb}.
They found that the energy gaps of AGNRs can be created by uniaxial strain driving the Fermi points between the confinement-induced discrete quasimomentum lines in the allowed electronic states\cite{linanores}.
For ZGNRs (zigzag graphene nanoribbons), the energy gaps can be slightly modulated by uniaxial strain, even when considering the polarization interactions at the edge states\cite{lunanores,linanores}.
\end{widetext}
\begin{figure}[htbp] 
%\centering 
\subfigure[]{ 
\label{Fig02a} 
\includegraphics[height=3.5cm,width=0.46\textwidth]{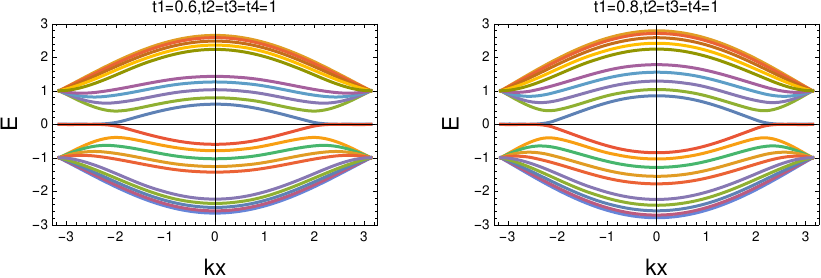}} 
%\caption{}
\subfigure[]{ 
\label{Fig02b} 
\includegraphics[height=3.5cm,width=0.46\textwidth]{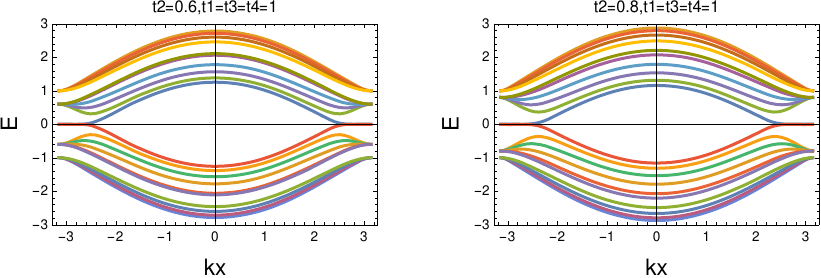}}
\subfigure[]{ 
\label{Fig02c} 
\includegraphics[height=3.5cm,width=0.46\textwidth]{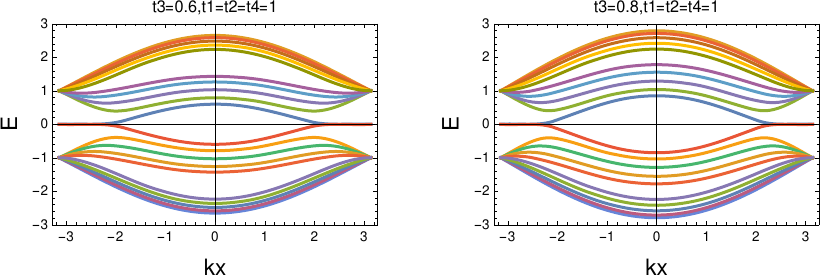}}
\subfigure[]{ 
\label{Fig02d} 
\includegraphics[height=3.5cm,width=0.46\textwidth]{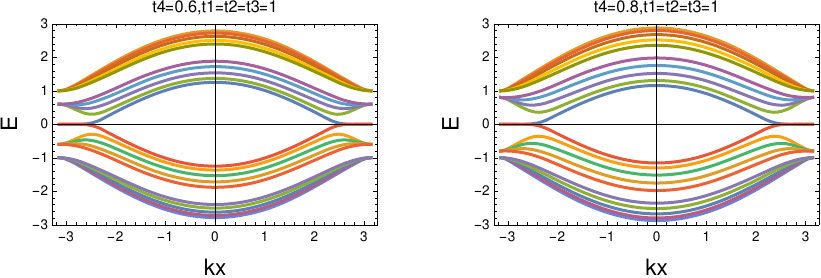}}
\caption {The energy minibands for the strained GNRs with altering single hopping term. (a) $t_1=0.8,0.6$;  (b)  $t_2=0.8,0.6$;  (c)  $t_3=0.8,0.6$; (d) $t_4=0.8,0.6$, the other hopping terms are assumed to be the pristine ones.}
\label{Fig02}
\end{figure}
\begin{figure}[htbp] 
\subfigure[]{ 
\label{Fig03a} 
\includegraphics[height=3.5cm,width=0.46\textwidth]{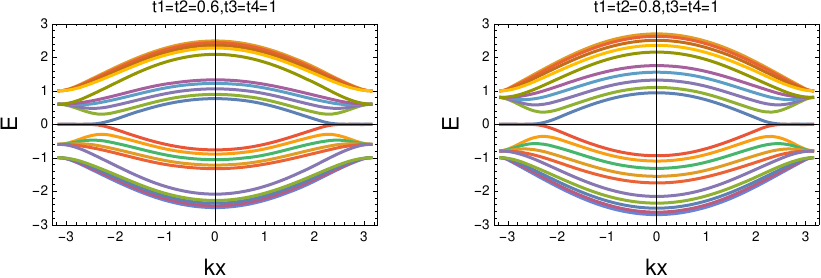}} 
\subfigure[]{ 
\label{Fig03b} 
\includegraphics[height=3.5cm,width=0.46\textwidth]{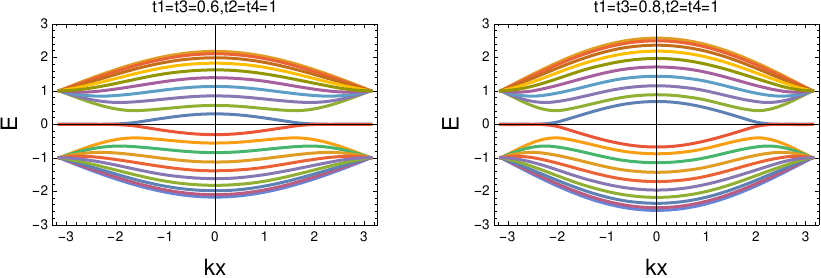}}
\subfigure[]{ 
\label{Fig03c} 
\includegraphics[height=3.5cm,width=0.46\textwidth]{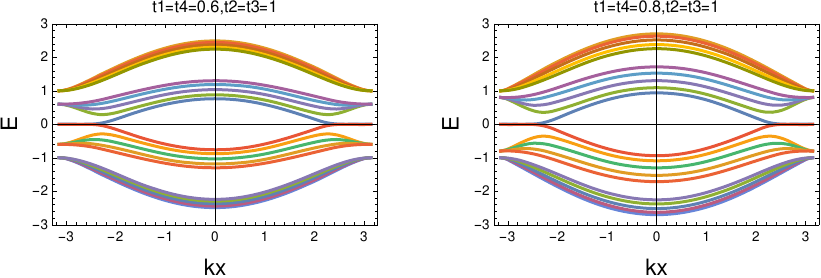}}
\subfigure[]{ 
\label{Fig03d} 
\includegraphics[height=3.5cm,width=0.46\textwidth]{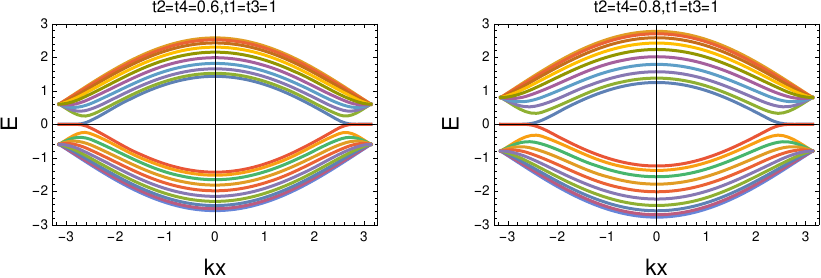}}
\caption {The energy minibands for the strained GNRs under various scenarios of hopping patterns 
(a) $t_1=t_2=0.6,0.8$, $t_3=t_4=1.0$; (b)  $t_1=t_3=0.6,0.8$, $t_2=t_4=1.0 $;
(c) $t_1=t_4=0.6,0.8$, $t_2=t_3=1.0 $; (d) $t_2=t_4=0.6,0.8$, $t_1=t_3=1.0 $.}
\label{Fig03}
\end{figure}
Firstly, let us focus on how the change of the hopping terms influence the energy minibands of GNRs. As illustrated in  Fig.\ref{Fig02a}-\ref{Fig02d} where only the single hopping term among the four hopping terms $(t_i,i=1,2,3,4)$ is altered. By comparing the figures in panels (a) and (c) with those in panels (b) and (d), it is manifestly clear that 
altering the hopping term $t_1,t_3$ alone (or $t_2,t_4$ alone) makes the minibands exhibit the similar shape. The reason behind this similarity lies in the fact that the hopping terms $t_1$  and $t_3$ enter the effective dispersive Hamiltonian (cf. Eq. \eqref{ham}) with the same multiplicative factor $q=2\cos(k_x a/2)$, as do the other two hopping terms $t_2$   and $t_4$, but with no multiplicative factor. The notable difference between Fig.\ref{Fig02a}\ref{Fig02c} and Fig.\ref{Fig02b},\ref{Fig02d} is primarily manifested in the fact that when changing $t_2$   or $t_4$ alone, new minibands emerge throughout the entire Brillouin Zone (BZ), including at the BZ boundary ($k_x = [-\pi, +\pi]$). In stark contrast, when changing $t_1$ or $t_3$ alone, the new momentum-resolved energy gaps only appear in the middle of the BZ and vanish at the BZ boundary. This phenomenon still stems from the multiplicative factor $q$, which decreases around the BZ boundary and even vanishes at BZ, thus, weakening the effect of hopping terms $t_1$ and $t_3$ at BZ, and failing to open a gap at BZ boundary.  It is noteworthy that the conduction and valence minibands in all figures exhibit symmetry about zero energy, signifying the preservation of electron-hole symmetry in the current model.

For more complex scenarios involving the alteration of two or more hopping terms\cite{jpc21}, the numerical results pertaining to the minibands are depicted in Fig.\ref{Fig03}. It is intriguing to observe that the conduction and valence minibands, influenced by the combined variations in hoppings ($t_1,t_2$)\cite{nature18a,nature18b,jpc21}, and ($t_1,t_4$),  not only widen the gap within BZ but also open a gap at the BZ boundary as shown in Fig.\ref{Fig03a}, \ref{Fig03c}.
However, contrary to intuition, the combined effect of ($t_1,t_3$) not only fails to widen the gap within the BZ as illustrated in Fig.\ref{Fig03b}, but also completely obliterated the existing gaps present when changing $t_1$ alone, as depicted in Fig.\ref{Fig02a}. Similarly, when ($t_2 ,t_4$) are simultaneously changed, the existing momentum-resolved gaps that emerge when $t_2$ or $t_4$ were altered individually, as shown in Fig.\ref{Fig02b}, \ref{Fig02d}, vanish entirely, including at the BZ boundary, as evident in Fig.\ref{Fig03d}.

Actually, the band edge profiles of minibands can be deduced from the Green's function, a tool commonly utilized to compute DOS and various electronic properties\cite{hhk}. In the context of tight-binding approximations, 
the HHK method, devised by Haine, Haydock, and Kelly, offers a potent solution to tackle real-space electronic structure. This Green's function can then be efficiently evaluated through recursive expansion via continued fractions.
Specifically, for infinite graphene GNRs whose unit cell contains four nonequivalent periodic carbon atoms, the continued fraction representation of the Green's function can be expressed as\cite{cresti08,vidarte_hhk}.
\begin{eqnarray}
\label{hhkgf}
\begin{aligned}
G\left(k_x,E\right)=\frac{1}{E-\frac{h_1^2}{E-\frac{h_2^2}{E-\frac{h_3^2}{E-h_4^2G(k_x,E)}}}}
\end{aligned}
\end{eqnarray}
Solving this continued fraction, one can obtain the analytical expression for Green's function. 
\begin{eqnarray}
%\label{eq28}
G\left(k_x,E\right) =\frac{\beta_1-\sqrt{\beta_2+\beta_1^2}}{ \beta_3}
\end{eqnarray}
where
\begin{eqnarray}
%\label{eq28}
\begin{aligned}
& \beta_1=\epsilon^4+\epsilon^2\sum_{j=1}^4(-1)^{1-\lfloor j/4\rfloor}h^2_j+\sum_{j=1}^2(-1)^{j-1}(h_jh_{j+2})^2 \\
& \beta_2=-4h^2_4\epsilon^2\prod_{j=1}^2\left(\epsilon-h_j^2-h^2_{j+1}\right)\\
& \beta_3=2h_4^2\epsilon\left(\epsilon^2-\sum_{j=1}^2 h_j^2\right)
\end{aligned}
\end{eqnarray}
where $\lfloor x \rfloor$ is a floor function. 
The discriminant for the continued fraction-form Green's function can be derived as follows,
\begin{eqnarray}
%\label{eq28}
\begin{aligned}
& D\left(k_x,E\right) =-4\epsilon^2\left(\epsilon^2-h_2^2 -h_3^2 \right)\left(h_4^2\epsilon^2-h_1^2h_4^2 -h_2^2h_4^2 \right) \\ &+\left(h_2^2h_4^2-h_1^2h_3^2+\epsilon^2\left(h_1^2+h_2^2+h_3^2-h_4^2-\epsilon^2\right)\right)^2
\end{aligned}
\end{eqnarray}
Solving above 8-th degree polynomial equations leads to the following solutions,
\begin{eqnarray}
\label{delimiting}
\begin{aligned}
& E_{\text{del.}}=\pm \sqrt{\frac{\gamma_1}{2}\pm \frac{1}{2}\left[\gamma_1^2-4\left(h_2^2h_4^2\pm 2h_1h_2h_3h_4+h_1^2h_3^2\right)\right]^{\frac{1}{2}}}\\ &
\end{aligned}
\end{eqnarray}
where $\gamma_1= h_2^2+h_4^2+\left(h_1^2+h_3^2\right)$.

\begin{figure}[htb!] 
\includegraphics[height=6cm,width=0.45\textwidth]{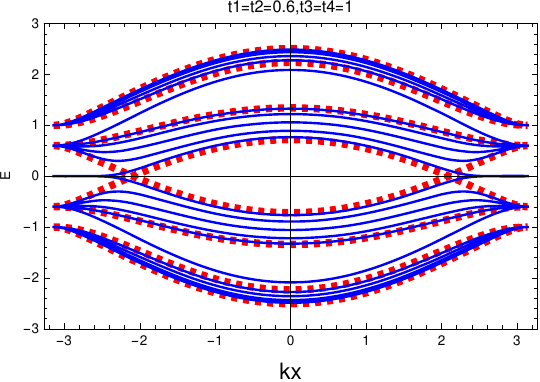} 
\caption {The energy minibands of finite GNRs ($N=4w$ and $N=20$) under the hopping pattern of $t_1=t_2=0.6$, $t_3=t_4=1.0$ versus the delimiting solution of the energy band edges computed from Green's function for infinite GNRs with the same hopping pattern, showing the remarkable match of the band edges.}
\label{Fig04}
\end{figure}

In Eq.\eqref{delimiting}, the eight roots yield delimiting curves that determine the energy band edges of infinite graphene nanoribbons (GNRs) whose unit cells comprise four nonequivalent carbon atoms in a periodic manner. The first symbol  "$\pm$" differentiates between the conduction (+) and valence bands at zero energy, while the other two  "$\pm$" symbols specify the miniband edge profiles within the conduction and valence bands, respectively. 
The delimiting curves arising from these eight solutions are depicted in Fig.\ref{Fig04} with red dashed lines, showcasing a remarkable match across almost all band edges. These
delimiting solutions offer an alternative means to validate the accuracy of the numerical results presented herein.

For practical computation of local density of states (LDOS), 
\begin{eqnarray}
\rho_i(E)=-\frac{1}{\pi}\int_{BZ} dk_x\underset{\epsilon\rightarrow 0}{\text{Im}}\text{Tr} \left[G_{i,i}(k_x,E+i\epsilon)\right]
\label{ldos}
\end{eqnarray} 
where $G_{i,i}(k_x,E)$ can be obtained as,
\begin{eqnarray}
\begin{aligned}
G_{i,i}(k_x,E)=\frac{1}{E-\frac{h_{[(i+3) \bmod 4]+1}^2 D_{i+1}}{D_{i-1}}-\frac{h_{[(i+2) \bmod 4]+1}^2 D_{i-1}^*}{D_i^*}}
\end{aligned}
\label{cfrac}
\end{eqnarray} 
where $D_i$ is the determinant of a matrix  by suppressing both the first $i$ rows and $i$ columns of the matrix $E-H$,
 while $D_i^*$ is the determinant of the matrix by extracting first $i$ rows and $i$ columns of the matrix $E-H$ simultaneously.

\begin{widetext}

\begin{figure}[htb!] 
\centering 
\subfigure[]{ 
\label{Fig05a} 
\includegraphics[height=3cm,width=0.7\textwidth]{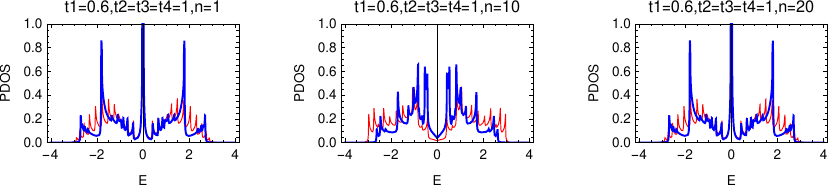}} 
\subfigure[]{ 
\label{Fig05b} 
\includegraphics[height=3cm,width=0.7\textwidth]{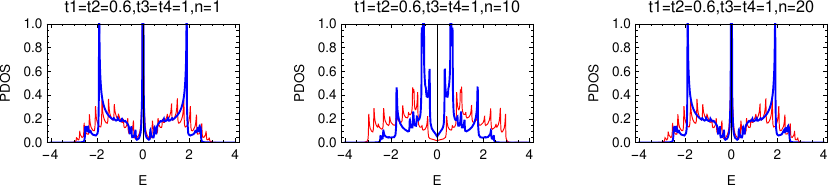}}
\subfigure[]{ 
\label{Fig05c} 
\includegraphics[height=3cm,width=0.8\textwidth]{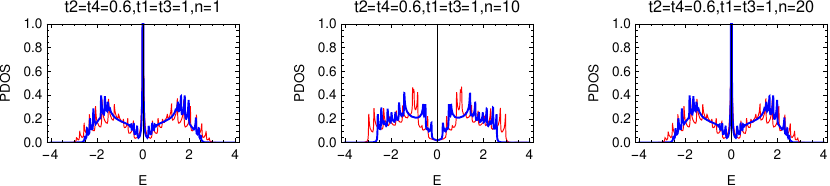}}
\caption {The local density of states for the strained GNRs under various scenarios (a) Pristine GNRs ($t_i=t$)}.
(b) $t_1=t_2=0.6$, $t_3=t_4=1.0$; (c)  $t_1=t_3=0.6$, $t_2=t_4=1.0 $;
(d) $t_2=t_4=0.6, t_1=t_3=1.0$.
\label{Fig05}
\end{figure}

\end{widetext}

  By employing  Eq.\eqref{ldos},\eqref{cfrac}, the numerical results of the LDOS  for the carbon atoms at site of $n=1,$ and $n=10, n=20$ are depicted in Fig.\ref{Fig05a}-\ref{Fig05c} where the thin red lines are for the prestine GNRs, and the blue thick lines are for strained GNRs.
The hopping patterns chosen herein leads to three types of different minibands, i.e., middle-hollowed-out, throughout-hollowed-out and no-hollowed-out structures as shown in Fig.\ref{Fig02a}, Fig.\ref{Fig03a}, and Fig.\ref{Fig03d} respectively.

The figures clearly show that regardless of the presence or absence of stress, the LDOS remains virtually identical for the first atom ($n=1$) and the last atom ($n=20$) within the unit cell of GNRs. 
This similarity is primarily attributed to the combined effects of the hard wall boundary condition and the inherent periodicity of the atomic arrangement within the unit cell. The energy range within which the LDOS is non-zero for pristine GNRs typically exceeds that of strained GNRs, owing to the reduction in the bandwidth of the energy bands under stress, as evident from the illustrations presented in Fig.\ref{Fig02},\ref{Fig03}.
In Fig.\ref{Fig05c}, the LDOS of both pristine and strained GNRs displays a remarkable similarity, which stems from the profound resemblance in their energy bands. This resemblance is mainly characterized by the absence of a hollowed-out structure, a common feature observed upon comparing Fig.\ref{Fig01b} and Fig.\ref{Fig03d}.

The most intriguing scenario can be seen in Fig.\ref{Fig05b}, compared to the regular peaks found in pristine GNRs, the peaks in strained graphene are sharper, yet their number decreases. This phenomenon is precisely associated with the through-hollowed-out minibands structures. 
The strain squeeze the energy band structure into two or more clustered minibands and leaving the momentum-reoslved gaps in-between, which increase the peaks' height, while concomitantly smear intrinsic peaks of pristine GNRs.

In order to find the eigenstates for the effective Hamiltonian $H$, a hard wall boundary condition for GNRs has been assumed: $c_{0}=c_{N+1}=0$.  By taking $C_1=(c_0,c_1)^T=(0,(\xi_1-\xi_2)/2i)^T$, which stems from the linear combination of the eigen-vectors for the total  transfer matrix ${\mathcal T}$ defined in Eq.\eqref{eq6}, the successive use of the total transfer matrix yields,
\begin{eqnarray}
\label{eq15}
\begin{pmatrix}
c^{(m)}_{4p}\\
c^{(m)}_{4p+1}
\end{pmatrix}=
\begin{pmatrix}
(-1)^{p+1}\sin p\theta_m \\
\frac{(-1)^p}{{\mathcal T}_{22}}\left({\mathcal T}_{11}\sin p\theta_m +\sin ( p+1 )\theta_m \right)
\end{pmatrix} 
\end{eqnarray}
where the index $m$ denotes the $m-$th solutions of Eq.\eqref{eqnum}, further use of Eq.\eqref{eq5trans} and adjusting the index lead to the following relation:
\begin{eqnarray}
\label{eq16}
\begin{aligned}
&  c^{(m)}_{4p}  =\left( -1\right)^{p+1}\sin p\theta_m;\\
& c^{(m)}_{4p-3} =\pm L_m\left( -1\right)^{p-1}\sin \left( w-p+1\right) \theta_m\\
& c^{(m)}_{4p-2} = K_m\left( -1\right)^{p-1}\left(t_1 t_4\sin (p-1)\theta_m - t_2 t_3\sin p\theta_m\right) \\
& c^{(m)}_{4p-1} = \pm U_m\left( -1\right)^{p-1}\left(t_1t_2\sin(w-p+1)\theta_m \right. \\ & \left. -t_3 t_4\sin \left( w-p\right) \theta_m\right)
\end{aligned}
\end{eqnarray}
where the symbol $\pm$ stands for the conduction and valence bands respectively,  $L_m,K_m,U_m$ are respectively defined as follows,
\begin{eqnarray}
\begin{aligned}
& L_m= \frac{t_2t_3t_4\sin\theta_m}{|E_m| \left(t_2t_3\sin\left(w+1\right)\theta_m -t_1t_4\sin w\theta_m \right)};\\
& K_m =\frac{\left(t_4\sin w\theta_m\right)} {{|q_m|}t_1\left(t_1t_4\sin w\theta_m -t_2t_3\sin\left(1+w\right)\theta_m \right)}; \\
& U_m =|E_m|\frac{\csc\theta_m\sin w\theta_m }{t_1t_2t_3|q_m|},
\end{aligned}
\end{eqnarray}

\begin{widetext}

\begin{figure}[htb!] 
\subfigure[]{ 
\label{Fig06a} 
\includegraphics[height=3cm,width=0.4\textwidth]{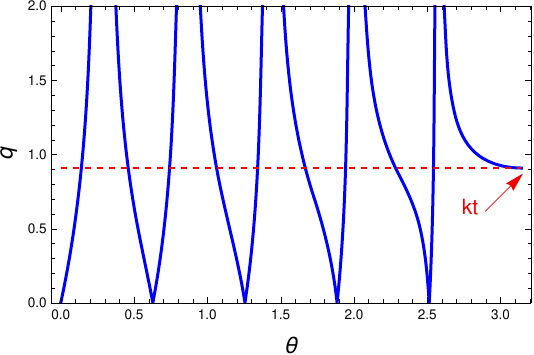}} 
\subfigure[]{ 
\label{Fig06b} 
\includegraphics[height=4cm,width=0.92\textwidth]{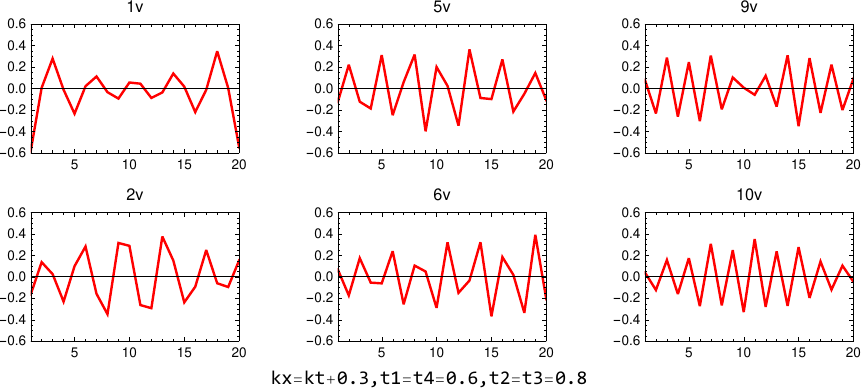}}
\subfigure[]{ 
\label{Fig06c} 
\includegraphics[height=4cm,width=0.92\textwidth]{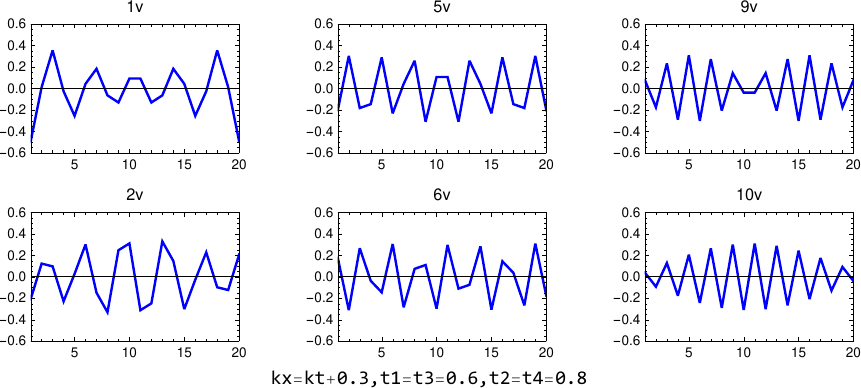}}
\subfigure[]{ 
\label{Fig06d} 
\includegraphics[height=4cm,width=0.92\textwidth]{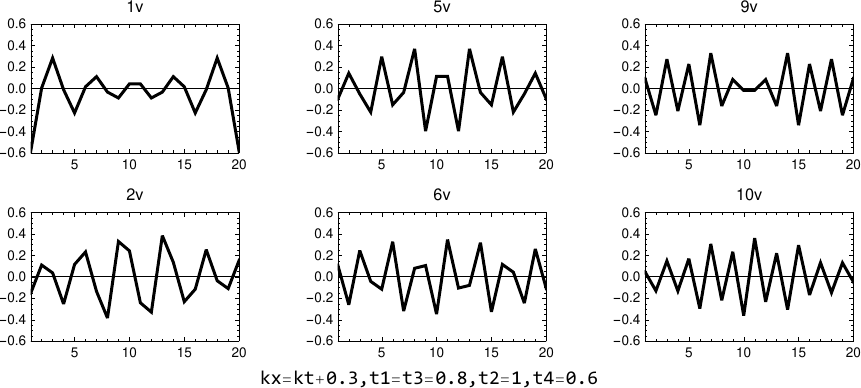}}
\caption {The edge states at $q=k_t+0.3$ and $k_t$ is the transition point. (a) the $q-\theta$, the transition point $k_t$ is indicated by red arrow, 
  (b) the wave function parity symmetry is broken $t_1=t_4=0.6$, $t_2=t_3=0.8$; (c)  the symmetry of wave fucntion parity is restored by setting $t_1=t_3=0.6$, $t_2=t_4=0.8$; 
	(d) the parity symmetry of the wavefunction remains preserved, when choosing $t_1=t_3=0.8$, $t_2=1.0, t_4=0.8$.}
\label{Fig06}
\end{figure}
\end{widetext}

To locate the edge states, one should first find $q-\theta$ relation by solving Eq.\eqref{qtheta}, and the numerical solution is depicted in Fig.\ref{Fig06a}. In this subfigure, the red arrow highlights the transition point $q_t$ where one real solution of $q(\theta_m)$  begins to vanish, persisting only in an analytical continuation as
 $q(\theta_m\rightarrow \pi+ i\theta_m)$. By adjusting the momentum $k_x$ to be around the transition point $q_t$, we can explore evolution of wavefunctions in bulk and edge modes\cite{saroka17}.

The wavefunctions are plotted at $k_x=k_t+0.3$  for different hopping term configurations in Fig.\ref{Fig06b} $t_1=t_4=0.6, t_2=t_3=0.8$, and Fig.\ref{Fig06c} $t_1=t_3=0.6, t_2=t_4=0.8$. 
In Fig.\ref{Fig06b}, it is evident that the wavefunctions corresponding to six valence minibands indexed by $m(v) $ ($m=$1,2,5,6,9,10) exhibit asymmetric structures across all 20 carbon atoms sites, indicating the disruption of parity symmetry  $(-1)^m$. Thus, the periodic strain in this case disrupts the parity symmetry, inherent in pristine GNRs (as mentioned in Ref.[\onlinecite{saroka17}]).

However, the wavefunction parity symmetry can be restored by setting $t_1=t_3$, and $t_2=t_4$, which has been shown in Fig.\ref{Fig06c}.
In this sub-panel, The wavefunction exhibits parity symmetry relative to the central point of the unit cell (specifically, the midpoint between the 10th and 11th atoms in this instance), and the miniband wavefunctions for the valence band undergo even or odd symmetry in an alternating manner 
 (i.e., for odd $m$, even symmetry shown (the upper panel of the Fig.\ref{Fig06c}), whereas for even $m$, odd symmetry observed (the lower panel of the Fig.\ref{Fig06c})) .  Finally, It is worth mentioning that in the $(1v)$ panels of both Fig.\ref{Fig06b},\ref{Fig06c}, the edge states can be seen. The wave functions of these edge states are mainly localized at the edges of the nanoribbons and decay rapidly towards the center of the nanoribbons.

When $t_1=t_3$, and $t_2=t_4$  are set, the model transforms into the $N=2w$ scenario, wherein the eigenvectors becomes,
\begin{eqnarray}
\label{w2wav}
\begin{aligned}
& c^{(m)}_{2 p}=(-1)^{p+1} \sin p \theta_m \\
& c^{(m)}_{2 p+1}= \pm(-1)^{p+1}(-1)^{m-1} \sin (p-w) \theta_m
\end{aligned}
\end{eqnarray}
where $\theta_m$ is the $m$-th solution of the following equations 
\begin{eqnarray}
\begin{aligned}
& -\frac{t_2}{q t_1} \sin (w \theta)+\sin ((w+1) \theta)=0 \\
& E= \pm \sqrt{q^2 t_1^2+t_2^2-2 q t_1 t_2 \cos\theta}~.
\end{aligned}
\end{eqnarray} 
From Fig.\ref{Fig06c}, it is evident that the recovery of parity symmetry of wavefunctions can be clearly observed, even if the GNRs is not pristine  under this circumstance.

As matter of fact, the parity of wavefunction can be resolved by introducing two operators: particle-hole symmetry operator $\Sigma^{z}$ and inversion operators $\Sigma^{x}$\cite{gusynin1}.
\begin{eqnarray}
\label{eq.a1}
\begin{aligned}
& \Sigma^{z}=\mathbb{I}_{nw}\otimes  \sigma_z \\
& \Sigma^{x}=\mathbb{J}_{nw} \otimes \sigma_x
\end{aligned}
\end{eqnarray}
where $\mathbb{I}_{nw}$,  $\mathbb{J}_{nw}$ are the identity matrix and row reversed identity matrix of size $2w$ respectively, $\sigma_x,\sigma_z$ are the Pauli matrices.
By using the periodic properties of Hamiltonian, it is easy to show 
\begin{eqnarray}
 \Sigma^{z} H\Sigma^{z}=-H 
\end{eqnarray}
indicating that when $|\psi\rangle $ is an eigenstate of Hamiltonian $H$,  $\Sigma^{z}| \psi  \rangle$ is also the eigenstate of $H$ but with opposite energy. 

When hopping terms $t_1=t_3$, 
we have further symmetry:
\begin{eqnarray}
\Sigma^{x} H\Sigma^{x}=H
\end{eqnarray}
indicating that the Hamiltonian $H$ and $\Sigma^x$ share the same eigenstates.
We can further prove that  $\Sigma^x   \Sigma^z|\psi \rangle = -  \Sigma^z|\psi\rangle$, therefore, the wavefuncton's parities are mutually opposite when their energies become opposite.

Without loss of generality, we assume $w$ is even number and hopping terms are positive. In order to use roots separation theorem\cite{phdxu} and  Min-Max principle\cite{golub,bookminmax},  after introducing a similarity transformation, $\mathbb{S}\Sigma^{x} H \mathbb{S}^{-1}$,  the following block matrix can be obtained as,
\begin{eqnarray}
\label{aplusa}
\mathbb{S}\Sigma^{x}H\mathbb{S}^{-1}=
\begin{bmatrix}
\mathbb{A} & 0\\
0 & \mathbb{A}
\end{bmatrix}
\end{eqnarray}
where $\mathbb{S}$  and $\mathbb{A}$  can be written as, 
\begin{eqnarray}
\label{mats}
 \mathbb{S}=\left(\begin{array}{c|c}
\delta_{2j-1,2j-1}  & \delta_{2j,2w+1-2j}\\\hline
\delta_{2j,2j} & \delta_{2j-1,2w+2-2j}
\end{array}\right)_{4w \times 4w} 
\end{eqnarray}
 and 
\begin{eqnarray}
\label{matA}
\mathbb{A}=\mathbb{H}_{2w}+\begin{pmatrix}
0 & \cdots & 0\\
\vdots & \ddots & \vdots\\
0 & \cdots & t_{4}
\end{pmatrix}_{2w \times 2w}, \text { $w$=~\text{even}}
\end{eqnarray}
where  the size of  $\mathbb{A}$ and four block matrices from $\mathbb{S}$ are $2w$, and  $j \leq w$, $\mathbb{H}_{2w}$ represents the 2w-sized submatrix by extracting the first $2w$ rows and columns from Hamiltonian $H$.

When suppressing the $2w-$th row and $2w-$th column of the Hamiltonian $H$, the new matrix  of size $4w-1$ can be arranged into the following block diagonal matrix, 
 \begin{eqnarray}
\tilde{\mathbb{V}}=
\begin{bmatrix}
\mathbb{V}_{2w-1} & 0\\
0 & \mathbb{V}_{2w}
\end{bmatrix}~.
\end{eqnarray}
The extended roots separation theorem\cite{phdxu} asserts the following inequality among the different groups of eigenvalues of the three matrices $\mathbb{V}_{2w-1}$, $\mathbb{V}_{2w}$, and Hamiltonian $H$:
\begin{eqnarray}
\begin{aligned}
 & E_{1}< \beta_{1}< E_{2}< \alpha_{1}< E_{3}< \beta_{2}< E_{4}< \alpha_{2}< E_{5} \\ & < \cdots \cdots <E_{4w-3}< \beta_{2w-1}< E_{4w-2}  < \alpha_{2w-1} \\ & < E_{4w-1}< \beta_{2w}< E_{4w}~,
\end{aligned}
\end{eqnarray}
where $\{\alpha_{i}\}$,  $\{\beta_{i}\}$   and $\{E_{i}\}$ are the eigenvalues of $\mathbb{V}_{2w-1}$, $\mathbb{V}_{2w}$, and Hamiltonian $H$ respectively which are arranged in an ascending order.  

Applying Min-Max principle to the matrix in Eq.\eqref{matA}, we obtain another inequality as,
\begin{eqnarray}
& \beta_1\leq e_1< \alpha_{1}< \beta_{2}\leq e_2< \alpha_{2}\cdots \cdots  \nonumber \\ & < \alpha_{2w-1}< \beta_{2w}\leq e_{2w}
\end{eqnarray}
where $\{e_{i}\}$ are the eigenvalues of $\mathbb{A}$ arranged in an ascending order.
 By comparing two inequalities, the eigenvalues of $\mathbb{A}$ can be inferred as: $\{E_{4w},E_{4w-2},E_{4w-4},\ldots E_{2}\}$ which are arranged herein in a descending order however.
 Due to particle-hole symmetry, we can re-index the above eigen-energies by merely counting the non-negative eigen-energies, then the eigenvalues of $\mathbb{A}$ can be written as:
 $\{E_{2w},-E_{2w-1},E_{2w-2},-E_{2w-3}\ldots ,-E_{1}\}$, therefore the eigenvalues of operator $\Sigma^x H$ can be cast into the following form:
\begin{eqnarray}
\mathbb{U}^{-1}\Sigma^{x}H\mathbb{U}=\begin{pmatrix}
E_{2w} & 0 & 0 & 0 & 0\\
0 & -E_{2w-1} & 0 & 0 & 0\\
0 & 0 & \ddots & 0 & 0\\
0 & 0 & 0 & -E_{2w-1} & 0\\
0 & 0 & 0 & 0 & E_{2w}
\end{pmatrix}
\end{eqnarray}
Since the inversion operator $\Sigma^{x}$ and Hamiltonian $H$  are commutative with each other, so they can be diagonalized simultaneously:
\begin{eqnarray}
\label{eseq}
\mathbb{U}^{-1}H\mathbb{U}=\begin{pmatrix}
E_{2w} & 0 & 0 & 0 & 0\\
0 & E_{2w-1} & 0 & 0 & 0\\
0 & 0 & \ddots & 0 & 0\\
0 & 0 & 0 & -E_{2w-1} & 0\\
0 & 0 & 0 & 0 & -E_{2w}
\end{pmatrix}~,
\end{eqnarray}
and 
\begin{eqnarray}
\label{parity}
\mathbb{U}^{-1}\Sigma^{x}\mathbb{U} =\begin{pmatrix}
1 & 0 & 0 & 0 & 0\\
0 & -1 & 0 & 0 & 0\\
0 & 0 & \ddots & 0 & 0\\
0 & 0 & 0 & 1 & 0\\
0 & 0 & 0 & 0 & -1
\end{pmatrix}_{4w\times 4w}~.
\end{eqnarray}
From above equation, it can be concluded that when $t_1=t_3$, the wavefunctions have definite parity, and the parities of the wavefunctions changes in an alternating manner 
per the eigen-energies sequence arranged in a descending/ascending order.
In order to corroborate the above result, the numerical computation of the wavefunctions has been given in Fig.\ref{Fig06d}, where the hopping term scenario are chosen to be $t_1=t_3=0.8$, and $t_2=1.0,t_4=0.6$.
From the figure, it can be clearly seen that the parities of wavefunctions precisely change in an alternating manner, i.e., the wavefunctions are either symmetric or anti-symmetric with respect to the center of unit cell of GNRs.

Lastly, we mention in passing that we have also proven the determinacy of parity for wavefunctions of periodically strained GNRs with more complex strain scenarios.

\section{Optical properties of GNRs due to strain}

To investigate the effect of strain on the optical properties of GNRs with zigzag edges, the usual approach is to compute the optical transition matrix within the gradient approximation: 
$M_{f,i}=\left\langle {f}|\bm{\hat{v}}\cdot \hat{\bm{e}}_{i}| i\right\rangle$, where $|i\rangle (|f\rangle)$ is the initial (final) state, and  $\hat{\bm{e}}_{i} $ is the incident electric field polarization unit vector and $\bm{\hat{v}}$ is the velocity operator defined as follows\cite{yanvoon_optexp}, 
\begin{eqnarray}
\label{eq26}
\hat{\bm v}=\frac{i}{\hbar }\left[\hat{H},\hat{\bm r}\right]=\frac1{\hbar }{\nabla_{\bm k} {H}({\bm k})}~.
\end{eqnarray}
When the final state $|f\rangle $ is $m-$th bands in conduction band denoted as $m(c)$ and initial state $|i \rangle$  is $n-th$ bands in valence band denoted as $n(v)$, then the velocity matrix element (VME) can be written as follows,   
\begin{widetext}

\begin{eqnarray}
\label{eq25}
\begin{aligned}
M_{n(c),m(v)} ={\sum }_{p=1}^{w}c_{4p-3(c)}^{\left(n\right)*}{\xi }_{4p-3(v)}^{\left(m\right)}+c_{4p-2(c)}^{\left(n\right)*}{\xi }_{4p-2(v)}^{\left(m\right)}+c_{4p-1(c)}^{\left(n\right)*}{\xi }_{4p-1(v)}^{\left(m\right)}+c_{4p(c)}^{\left(n\right)*}{\xi }_{4p(v)}^{\left(m\right)}
\end{aligned}
\end{eqnarray}
where $|{\xi }_{j(v)}^{(m)}\rangle~ (j=4p-3,4p-2,4p-1,4p)$,
\begin{eqnarray}
\label{eq26a}
\left|\xi_{j(v)}^{(m)}\right\rangle=\frac{a}{\hbar }\frac{\partial H\left(k\right)}{\partial k}\left |c_{j(v)}^{\left(m\right)}\right\rangle 
\end{eqnarray}
Then $|{\xi }_{j(v)}^{(m)}\rangle$ can be obtained via Eq.\eqref{eq16},
\begin{eqnarray}
\label{xi4cycle}
\begin{aligned}
& \xi^{(m)}_{4p-3} =-\frac{a}{\hbar }t_1 K_m\left[ -1\right)^{p-1}\left(t_1t_4\sin \left[(p-1)\theta_m\right]-t_2t_3\sin \left(p\theta_m\right)\right] \\
& \xi^{(m)}_{4p-2}=\mp\frac{a}{\hbar }t_1 L_m\left( -1\right)^{p-1}\sin \left[\left(w-p+1\right) \theta_m\right]\\
& \xi^{(m)}_{4p-1}=-\frac{a}{\hbar }{t_3}\left( -1\right)^{p+1}\sin \left(p\theta_m\right)\\
&\xi^{(m)}_{4p} =\mp\frac{a}{\hbar }t_3 U_m\left( -1\right)^{p-1}\left [t_3t_4\sin\left[ \left( w-p\right) \theta_m\right]- t_1t_2\sin\left[(w-p+1)\theta_m\right]\right]
\end{aligned}
\end{eqnarray}

Inserting Eq.\eqref{eq16} and Eq.\eqref{xi4cycle} into Eq.\eqref{eq25}, the VME can be written as,
\begin{eqnarray}
\label{eq27}
\begin{aligned}
M_{\left(c\right)n,\left({v}\right)m} &=\frac{{a}}{{\hbar }}N_n N_m \sin \left(\frac{k_x a}{2}\right)\left(-t^2_1 t_4  L_n  K_m S_{nm}^1+t_1 t_2 t_3 L_n  K_m S_{nm}^2+t^2_1 t_4 L_m  K_n S_{nm}^1 \right. \\
& -t_1 t_2 t_3 L_m  K_n S_{nm}^2+ t_1 t_2t_3  U_n S_{nm}^2-t^2_3 t_4 U_n S_{nm}^1 \\
& \left. -t_1 t_2 t_3  U_m S_{nm}^2+ t^2_3 t_4  U_m S_{nm}^1\right)
\end{aligned}
\end{eqnarray}
where $N_{n}$,$N_{m}$ are the normalization coefficients for the eigenvectors and defined in the following Eq.\eqref{eq20}, 
\begin{eqnarray}
\label{eq20}
N_m=\frac{1} {\sqrt{{\sum}_{p=1}^{w}c_{4p-3}^{\left(m\right)*}c_{4p-3}^{\left(m\right)}+c_{4p-2}^{\left(m\right)*}c_{4p-2}^{\left(m\right)}+c_{4p-1}^{\left(m\right)*}c_{4p-1}^{\left(m\right)}+c_{4p}^{\left(m\right)*}c_{4p}^{\left(m\right)}}}~,
\end{eqnarray}
\end{widetext}
while $S_{nm}^1$, $S_{nm}^2$ are expressed as,
\begin{eqnarray}
\label{eq28}
\begin{aligned}
S_{n,m}^1  & =\frac{\sin \left(w\theta_m \right)\sin \theta_n-\sin \theta_m\sin \left(w\theta_n\right)}{2\cos \theta_m-2\cos \theta_n} \\
S_{n,m}^2  & =\frac{\sin \left[(w+1)\theta_m\right]\sin \theta_n-\sin \theta_m\sin \left[(w+1)\theta_n\right]}{2\cos \theta_m-2\cos \theta_n}~.
\end{aligned}
\end{eqnarray}
Despite the complexity of Eq.\eqref{eq27}, it is evident that the interband optical matrix $M_{m(c),n(v)}$  vanishes (as the terms within the bracket in Eq. \eqref{eq27} cancel each other out) when the conduction and valence bands share the same miniband index, i.e., $m=n$ or $\Delta J=m-n=0$. This indicates that interband optical transitions are prohibited between conduction and valence minibands with identical indices (distinguished by $\theta_m$). 
Additionally, it holds true that  $M_{m(c),n(v)}= -M_{n(c),m(v)}$ , which can be confirmed by exchanging $m$ and $n$.

The numerical calculation of the VME has been performed and presented in Fig.\ref{Fig07}. The top panel of the figure (Fig.7(a)) depicts the pristine case, while the other three panels show the strained scenarios. Notably, in all strained cases, the VME decrease across all possible interband transitions from the first valence band to the conduction band $n$ (denoted as $1(v)\rightarrow n(c)$). This attenuation primarily stems from the flattening of the minibands under strain. As the velocity operator is directly proportional to the first derivative of $E(k)$ with respect to $k$, the diminished slope of the flattened minibands reduces this derivative, resulting in smaller VME values.

In pristine GNRs, the interband transition adheres to the selection rule of  $\Delta J=|m-n|=\text{odd number}$\cite{sasaki,mflin11,saroka17}. The top panel of Fig.\ref{Fig07} confirms this rule by illustrating the anticipated interband transition pattern, specifically showing transitions between $1v\rightarrow 2c, 1v\rightarrow 4c, 1v\rightarrow 6c, 1v\rightarrow 8c, 1v\rightarrow 10c$, through the finite values of VME. However, a dramatic deviation from this selection rule is observed when the hopping term pattern shifts to $t_1=t_4=0.6$ and $t_2=t_3=1$, as illustrated in the first middle panel of the figure (Fig.7(b)). Remarkably, this alteration allows for almost all possible transitions (some transition amplitudes are small, but they cannot be considered negligible), except for the forbidden $1v\rightarrow 1c$ transition predicted by Eq.\eqref{eq27}, thereby entirely violating the interband optical transition selection rule for pristine GNRs. When the hopping pattern changes to $t_1=t_3=0.6$, and $t_2=t_4=1$, as shown in the second middle panel of the figure (Fig.7(c)), an intriguing phenomenon emerges: the interband optical transitions occur between  $1v\rightarrow 2c, 1v\rightarrow 4c, 1v\rightarrow 6c, 1v\rightarrow 8c, 1v\rightarrow 10c$, showing the restoration of the optical selection rule of $\Delta J=|m-n|=\text{odd number}$. More interesting phenomena is the VME profiles shown in the bottom panel of Fig.\ref{Fig07} where the same hopping pattern as that of Fig.\ref{Fig06d} is employed and only $t_1=t_3$ is assumed. In this bottom panel, the optical selection as that of prisitine GNRs is still retained, further demonstrating that our proof on the wavefunction parity is correct.
\begin{widetext}

\begin{figure}[htb!] 
\centering  
\includegraphics[height=16cm,width=0.7\textwidth]{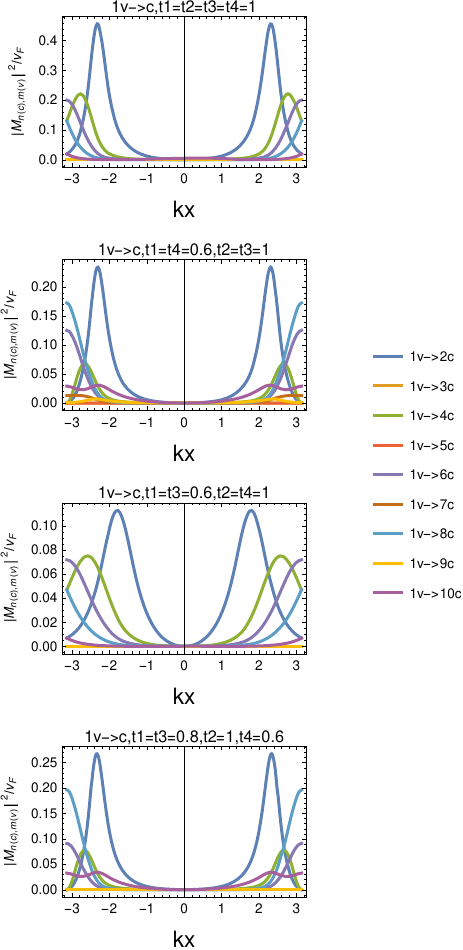} 
\caption {The interband VME for pristine and strained GNRs, the legends on the right hand of plots represent the optical transition $1v\rightarrow 2c, 1v\rightarrow 3c, \cdots  1v\rightarrow 9c, 1v\rightarrow 10c$ respectively.
  (a) the pristine GNRs: $t_1=t_2=t_3=t_4$; (b)  the strained case of $t_1=t_4=0.6$, $t_2=t_3=1.0$, the breaking of optical selection rule for pristine GNRs due to wavefunction parity can be clearly seen; (c) the strained case of $t_1=t_3=0.6$, $t_2=t_4=1.0$,  showing the restoration of the interband optical selection rule but with different VME values from pristine GNRs in (a); (d) the strained case of $t_1=t_3=0.8$, $t_2=1.0, t_4=0.8$, the same selecton rule as the pristine is still retained. }
\label{Fig07}
\end{figure}

\begin{figure}[htb!]  
\includegraphics[height=12cm,width=0.7\textwidth]{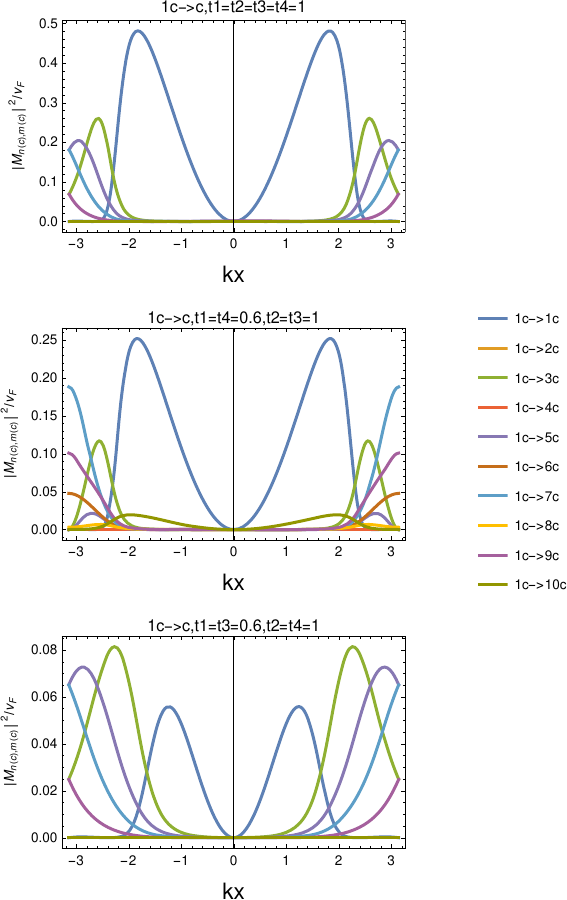} 
\caption {The intraband velocity matrix elements for pristine and strained GNRs, the legends of $1,2,\cdots, 8,9$ on the right hand of plots represent the optical transition $1c\rightarrow 1c$, $1c\rightarrow 3c, 1c\rightarrow 5c, 1c\rightarrow 7c, 1c\rightarrow 9c$ respectively.
  (a) the pristine GNRs $t_1=t_2=t_3=t_4$; (b)  the strained case of $t_1=t_4=0.6$, $t_2=t_3=1.0$, the breaking of optical selection rule for pristine GNRs due to wavefunction parity can be clearly seen. (c) the strained case of $t_1=t_3=0.6$, $t_2=t_4=1.0$,  the restoration of the interband optical selection rule.}
\label{Fig08}
\end{figure}

\end{widetext}

Regarding the selection rule for the interband VME in the context of a $N=2w$ strain-periodic system, the interband VME can indeed be analytically derived. The component vectors involved in this derivation are outlined below.
\begin{eqnarray}
\label{2periodeigen}
\begin{aligned}
& \xi_{2 p-1}^{(m)}=-\frac{t_1 a}{\hbar} \sin \left(\frac{k_x a}{2}\right) \sin (p \theta_m) \\
& \xi_{2 p}^{(m)}= \pm \frac{t_1 a}{\hbar} \sin \left(\frac{k_x a}{2}\right)(-1)^{m-1} \sin [(w+1-p) \theta_m]
\end{aligned}
\end{eqnarray}
From these vectors, the VME for interband optical transition can be straightforward derived as follows,
\begin{eqnarray}
\label{2cycle}
\begin{aligned}
& M_{n(c), m(v)}=\frac{t_1 a}{\hbar} \sin \left(\frac{k_xa}{2}\right) N_n N_m\left[(-1)^n-(-1)^m\right] S_{n, m} \\
&S_{n, m}=\frac{\sin\theta_m\sin[(w+1)\theta_n]-\sin[(w+1)\theta_m]\sin\theta_n }{2(\cos\theta_n-\cos\theta_m)}
\end{aligned}
\end{eqnarray}
where $S_{n, m}$ retains the structure akin to Ref. [\onlinecite{saroka17}], the interband optical selection remains consistent with that of pristine GNRs, even under the influence of strain. Notably, in the aforementioned Eq.\eqref{2cycle}, $t_2$ implicitly contributes to the expression via $\theta_n$, suggesting that the condition for optical selection rule of pristine GNRs can be relaxed if the hopping patterns can be mapped onto an effective dimerized chain\cite{naumis14}. Upon comparing Eq.\eqref{2periodeigen} for $N=2w$ with Eq.\eqref{xi4cycle} for $N=4w$, a significant observation emerges: the wavefunction parity factor $\left( -1\right)^n$  vanishes in the latter scenario. This disappearance of wavefunction parity fundamentally distinguishes the interband optical transition selection rules between the $N=2w$ and $N=4w$ cases.

\begin{widetext}
As depicted in the energy miniband plots presented in Fig.\ref{Fig02},\ref{Fig03}, a judicious choice of hopping term patterns can introduce non-negligible momentum resolved gaps within both the conduction and valence bands. We now proceed to calculate the intraband VME for optical transitions within the conduction band $n(c)\rightarrow m(c)$ and within the valence band  $n(v)\rightarrow m(v)$ and the intraband VME can be calculated as,
\begin{eqnarray}
\label{intraopt}
\begin{aligned}
M_{n (c),m (c)} &=\frac{a}{\hbar }N_n N_m\sin \left(\frac{k_x a}2\right)\left(-t^2_1 t_4 L_n  K_m S_{nm}^1+t_1 t_2 t_3 L_n  K_m S_{nm}^2 \right. \\ 
& -t^2_1 t_4 L_m  K_n S_{nm}^1+t_1 t_2 t_3 L_m K_n S_{nm}^2+ t_1 t_2t_3 U_n S_{nm}^2-t^2_3 t_4  U_n S_{nm}^1 \\
& \left. +t_1 t_2 t_3  U_m S_{nm}^2-t^2_3 t_4  U_m S_{nm}^1\right)\\
M_{n (v),m(v)} &=\frac{a}{\hbar }N_n N_m\sin \left(\frac{k_x a}2\right)\left(t^2_1 t_4 L_n  K_m S_{nm}^1 - t_1 t_2 t_3 L_n K_m S_{nm}^2 \right. \\ & 
\left. +t^2_1 t_4 L_m K_n S_{nm}^1 - t_1 t_2 t_3 L_m K_n S_{nm}^2 - t_1 t_2 t_3 U_n S_{nm}^2+t^2_3 t_4 U_n S_{nm}^1 \right. \\
& \left.  -t_1 t_2 t_3 U_m S_{nm}^2+t^2_3 t_4 U_m S_{nm}^1\right)
\end{aligned}
\end{eqnarray}
\end{widetext} 
 In contrast to the optical interband transition, which is prohibited under the condition of $\Delta J=0$,  the intraband transtion still persists for strained GNRs under the condition of $\Delta J=0$, which can be deduced from intraband VME in Eq.\eqref{intraopt} by careful evaluation of the limiting behavior ($\theta_m \rightarrow\theta_n$) with L'Hospital rule:
\begin{eqnarray}
%\label{eq28}
\begin{aligned}
& S_{m,m}^1=\frac{1}{2}\left[-w\cos\left(w\theta_m\right)+\cot\theta_m\sin\left(w\theta_m\right)\right]\\
& S_{m,m}^2=\frac{1}{4}\csc\theta_m \left[\left(w+2\right)\sin\left(w\theta_m\right)-w\sin\left[\left(w+2\right)\theta_m\right]\right]~,
\end{aligned}
\end{eqnarray}
indicating that the intraband optical transition is allowed since the terms on the right hand side of Eq.\eqref{intraopt} fail to cancel out, unlike in the case of the interband optical transition.

The numerical computations of the VME ($M_{n(c),m(v)}$)  have been undertaken to demonstrate that the $N=4w$ scenario fundamentally alters the selection rule of the pristine GNRs.
For the pristine GNRs, the intraband optical transition selection rule is proportional to $(-1)^n+(-1)^m$, where $m,n$ are the index of conduction minibands. From upper panel of Fig.\ref{Fig08}, it can be clearly seen that the four VME curves corresponding respectively to $1c\rightarrow 1c$, $1c\rightarrow 3c$, 
 $1c\rightarrow 5c,$ $1c\rightarrow 7c,$ $1c\rightarrow 9c$ attain a finite value, while other VME vanishes, which truly obeys the selection rule $\Delta|J|=\text{even number}$.
\begin{figure}[htb!] 
%\centering 
\subfigure[]{ 
\label{Fig09a} 
\includegraphics[height=4cm,width=0.46\textwidth]{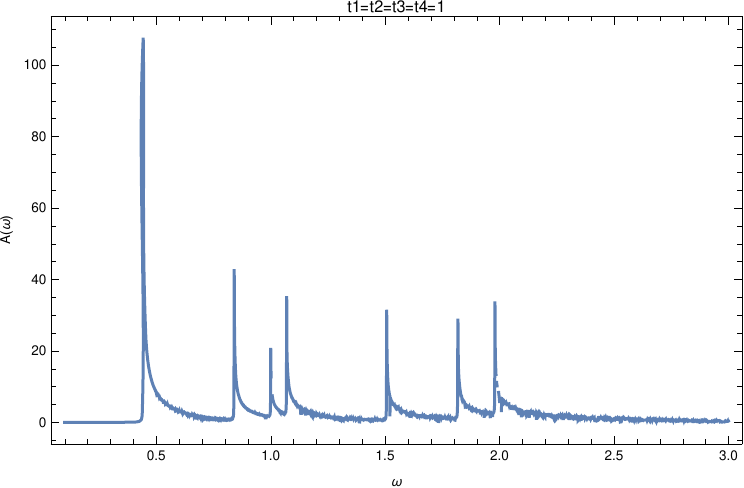}} 
%\caption{}
\subfigure[]{ 
\label{Fig09b} 
\includegraphics[height=4cm,width=0.46\textwidth]{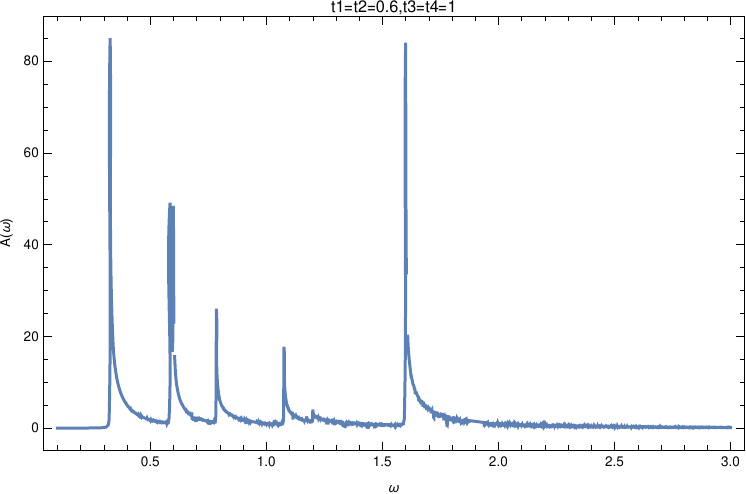}}
\subfigure[]{ 
\label{Fig09c} 
\includegraphics[height=4cm,width=0.46\textwidth]{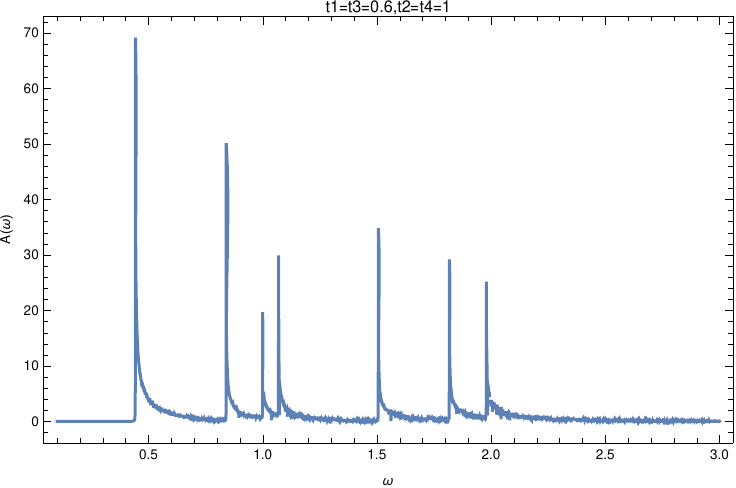}}
\subfigure[]{ 
\label{Fig09d} 
\includegraphics[height=4cm,width=0.46\textwidth]{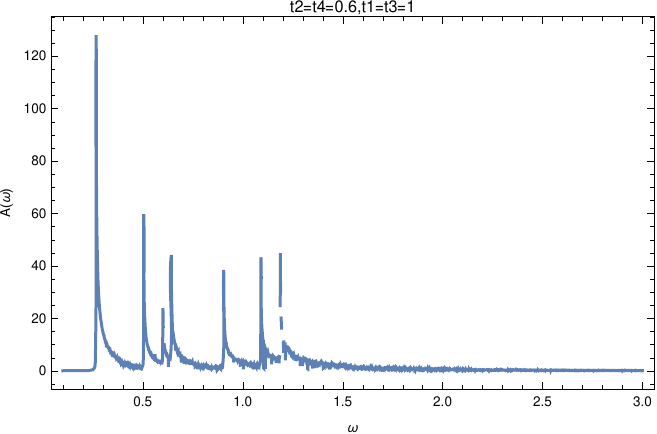}}
\caption {The absorption profiles for the strained GNRs at $0~ K$ (a) Pristine GNRs ($t_i=t$)}.
(b) $t_1=t_2=0.6$, $t_3=t_4=1.0$; (c)  $t_1=t_3=0.6$, $t_2=t_4=1.0 $;
(d) $t_2=t_4=0.6, t_1=t_3=1.0$.
\label{Fig09}
\end{figure}
However, as depicted in middle panel of Fig.\ref{Fig08}, when the hopping pattern is altered to $t_1=t_4=0.6$ and $t_2=t_3=1$,
 a stark deviation occurs. Almost all possible intraband transitions from  $1(c)\rightarrow n(c) $ (where $n\neq 1$) virtually attain finite values, thereby completely disregarding the $\Delta|J|=\text{even number}$ observed in pristine GNRs.

If we adjust the hopping parameters to $t_1=t_3=0.6$ and $t_2=t_4=1$, the intraband optical transition selection rule for pristine GNRs can indeed be reinstated. As evidenced in the lower panel of Fig.\ref{Fig08}, the numerical outcomes for this scenario showcase the intraband optical transitions exclusively confined to the transitions $1c\rightarrow 1c,$ $1c\rightarrow 3c,$  $1c\rightarrow 5c,$ $1c\rightarrow 7c,$  $1c\rightarrow 9c$, which is virtually the same as the pristine GNRs case.
The intraband optical transition of valence to valence minibands has also numerically been computed, and the numerical results bear the similarity to that of conduction minibands to conduction minibands transition due to electron-hole symmetry, thus, the results is omitted for brevity.

Let us now proceed to discuss the interband optical absorption, which is defined as follows:
\begin{eqnarray}
\begin{aligned}
 A(\omega) \propto & \lim_{\Gamma\rightarrow 0} \sum_{n, m}\int_{BZ}\frac{dk}{2\pi} \operatorname{Im}\left[\frac{f\left(E_{c,m}(k)\right)-f\left(E_{v,n}(k)\right)}{E_{v,n}(k)-E_{c,m}(k)-\omega-i \Gamma}\right] \\ & \times \frac{\left|M_{(c)n, (v)m}(k)\right|^2}{\omega},
\end{aligned}
\end{eqnarray}
where $f(E_{s,m}(k))$ is the Fermi-Dirac distribution function of $m$-th miniband from $s$-band (valence or conduction) at energy of $E_{s,m}(k)$.
The numerical findings have been visually presented in Fig.\ref{Fig09}, showcasing the variation in absorption characteristics when considering different hopping-changing combinations $((t_1,t_3),(t_1,t_2),(t_1,t_4))$ as compared to those of pristine GNRs. 
A common trend observed across sub-panels Fig.\ref{Fig09b}-\ref{Fig09d}  associated with strained GNRs is that the absorption diminishes more rapidly with increasing $\omega$ compared to the pristine counterpart Fig.\ref{Fig09a}. This expedited decay can be attributed to the shrinkage of the miniband width. 
Specifically, as $t_1$ and $t_3$  are simultaneously reduced, a previously mentioned effect of miniband flattening in both the conduction and valence bands emerges, leading to a pronounced enhancement in the joint density of states (JDOS). This enhancement is intimately linked to the inverse of the momentum derivative of the energy difference between the conduction and valence minibands ($1/\partial_{k_x}(E_c(k_x)-E_v(k_x))|_{E_c-E_v=\hbar\omega}$), a phenomenon akin to the van Hove singularity\cite{jdos16}. Consequently, as depicted in Fig.\ref{Fig09c}, the absorption peaks exhibit a more pronounced sharpness and narrowness compared to those observed in the pristine GNRs (Fig.\ref{Fig09a}). This underscores the significance of the mutual influence between the hopping parameters and the resulting miniband structure on the optical absorption properties of strained GNRs.

The absorption spectrum presented in Fig.\ref{Fig09b} specifically addresses the case where both $t_1$ and $t_2$ are modified, showing a decrease in the number of sharp resonant peaks from 7 to 5. This decrease can be attributed to the appearance of momentum-resolved gaps in both conduction and valence bands. The presence of these momentum-resolved energy gaps carves out the complete and regular energy bands into a hollowed-out structure, which in turn reduces both the probability of interband optical transitions and the number of peaks in the JDOS as LDOS\cite{jdos16}, thereby, leading to a diminished count of peaks
 In essence, the alteration of both hopping parameters ($t_1$ and $t_2$)  profoundly remodels the energy landscape, consequently modifying the optical absorption characteristics, manifested by a noticeable reduction in the number of peaks.

 Chang et al calculated the zigzag ribbon under the influence of the strain, and found that the absorption peaks started at $\omega\approx 0.5 t_0$, and the became featureless beyond $\omega\approx 2t_0$, which bears similar profiles as 
 Fig.\ref{Fig09a} and Fig.\ref{Fig09c}. While in Fig.\ref{Fig09b} and Fig.\ref{Fig09d}, there exists the obvious redshift, resulting from the strain-induced changes of the energy mionibands, see Fig.\ref{Fig03a} and 
 Fig.\ref{Fig03d}. From Fig.\ref{Fig03a}, it can be clearly seen that two minibands split the energy band gap of the prisitine GNRs into two smaller gaps, thereby effectively narrowing the interval of the first energy gap.
 While in Fig.\ref{Fig03d}, the candid reduction of the energy gap leads to the redshift of the interband transition.
Jia and Gao numerically investigated  the the effect of strain on the optical interband transition of $E_{11}$ and $E_{22}$  using the nonorthogonal tight-binding 
model\cite{pssbjia}. Their numerical method employs structural optimization, which involves intricate structural changes, unlike our assumption herein, which leads to a consistent reduction in hopping integrals. 

\section{Concluding remarks}
To summarize, we have delved into the impact of strain on the electronic and optical characteristics of GNRs utilizing an effective dispersive tri-diagonal Hamiltonian. 
The unit cell for the GNRs consists of periodic four nonequivalent carbon atoms whose nearest hoppings $(t_i, i=1,2,3,4)$ are designed to be varied from the pristine GNRs and the number of zigzag chain containing the four atoms in a periodic manner is $w$ (with $N=4w$ number of
carbon atoms in the unit cell). The analytical expression for the electronic and optical properties of GNRs has been obtained by leveraging the transfer matrix in a tractable manner and various strained GNR scenarios characterized by distinct hopping patterns have been explored.

Intriguingly, in scenarios where hopping decreases due to stress-induced elongation of carbon bonds, we observe that certain hopping patterns can introduce momentum-resolved gaps within both the conduction and valence bands, making the GNRs energy bands look like hollowed-out structures, while others simply lead to the expected shrinkage of band width. The validity of our numerical results for the minibands is corroborated by the remarkable alignment of the miniband profiles with the delimiting solutions of the discriminant derived from the continued fraction form of the Green's function within the framework of the Heine-Haydock-Kelly (HHK) methods\cite{hhk}.

The LDOS for carbon atoms at different sites within the unit cell has been calculated and contrasted across various hopping patterns, with the evolution of minibands serving as the basis for comparison. The effect of strain on the interband and intraband optical transitions has been conducted by leveraging VMEs. Notably, a stark contrast emerges in the selection rules compared to those inherent in pristine GNRs, where interband transitions between minibands with same (different) parity index are forbidden (allowed), whereas intraband transitions just obey the opposite rule\cite{mflin11,saroka17}. Deeper numerical exploration into wavefunction parity symmetry breaking offers analytical insights into the alterations in selection rules.

As matter of fact, both the optical selection rules and wavefuncion parity symmetry can be restored by setting the hopping pattern to be $t_1=t_3$, and $t_2=t_4$, relaxing the pristine GNRs (i.e., $t_i=t$) while recovering the optical selection rules and parity symmetry. By employing roots separation theorem and Min-Max principle, we have further proved that the parity of wavefunction still retained if  relaxing the condition to  $t_1=t_3$ only. 
In addition, the interband absorption has also been computed for different hopping patterns, The resulting absorption curves are the competition mechanisms
between VMEs which is reduced due to the decrease of the gradient of energy bands and the JDOS which is increased due to the shrinkage  of the energy bands. 

Remarkably, both the optical selection rules and wavefunction parity symmetry can be reinstated by adjusting the hopping pattern to $t_1=t_3$, and $t_2=t_4$, i.e., in a dimerized configuration\cite{naumis14}, relaxing the optical selection rule condition for pristine GNR configuration. Furthermore, we have computed interband absorption for diverse hopping patterns, revealing intricate competition mechanisms between VME diminished by reduced energy band gradients and JDOS enhanced by energy band shrinkage.
These findings hold promise for future investigations exploring a broader range of hopping patterns, potentially achieved through strain or topological band engineering\cite{nature18a,nature18b,jpc21}, for example, by increasing the number of non-equivalent carbon atoms within the GNRs unit cell.

%\nocite{*}

\end{document}